%% file: main.tex
\documentclass[sigconf]{acmart}

\setcopyright{none} 

\renewcommand\footnotetextcopyrightpermission[1]{}

\usepackage{xspace}
\def \sysname{\texttt{ASTM}\xspace}

\usepackage{pifont}
\usepackage{multirow}
\usepackage{adjustbox}

\usepackage{subfigure}

\usepackage[linesnumbered,ruled,vlined,noend]{algorithm2e}

\begin{document}


\title{Automation Slicing and Testing for in-App Deep Learning Models}




\author{Hao Wu}
\affiliation{\institution{National Key Laboratory for Novel Software Technology, Nanjing University}}

\author{Yuhang Gong}
\affiliation{\institution{National Key Laboratory for Novel Software Technology, Nanjing University}}

\author{Xiaopeng Ke}
\affiliation{\institution{National Key Laboratory for Novel Software Technology, Nanjing University}}

\author{Hanzhong Liang}
\affiliation{\institution{National Key Laboratory for Novel Software Technology, Nanjing University}}

\author{Minghao Li}
\affiliation{\institution{Harvard University}}

\author{Fengyuan Xu}
\authornote{Corresponding author. Email: fengyuan.xu@nju.edu.cn}
\affiliation{\institution{National Key Laboratory for Novel Software Technology, Nanjing University}}

\author{Yunxin Liu}
\affiliation{\institution{Institute for AI Industry Research (AIR), Tsinghua University}}

\author{Sheng Zhong}
\affiliation{\institution{National Key Laboratory for Novel Software Technology, Nanjing University}}

\begin{abstract}
Intelligent Apps (iApps), equipped with in-App deep learning (DL) models, are emerging to offer stable DL inference services. However, App marketplaces have trouble auto testing iApps because the in-App model is black-box and couples with ordinary codes. In this work, we propose an automated tool, \sysname, which can enable large-scale testing of in-App models. \sysname takes as input an iApps, and the outputs can replace the in-App model as the test object. \sysname proposes two reconstruction techniques to translate the in-App model to a backpropagation-enabled version and reconstruct the IO processing code for DL inference. With the \sysname's help, we perform a large-scale study on the robustness of 100 unique commercial in-App models and find that 56\% of in-App models are vulnerable to robustness issues in our context. \sysname also detects physical attacks against three representative iApps that may cause economic losses and security issues.
\end{abstract}

\maketitle

\input{intro_v2} 
\input{background}

\input{overview}

\input{code_recon}

\input{model_recon}
\input{expr}

\input{conclusion}

\bibliographystyle{ACM-Reference-Format}
\bibliography{main}

\end{document}

%% file: intro_v2.tex
\section{Introduction}
\label{sec:intro}

Deep learning (DL) technologies have significantly advanced many fields critical to mobile applications, such as image
understanding, speech recognition, and text translation~\cite{voulodimos2018deep,kamath2019deep, singh2017machine}.
Besides, a lot of research efforts have been put into optimizations of DL latency and
efficiency~\cite{liang2021pruning,berthelier2021deep,gou2021knowledge,menghani2021efficient}, paving the path towards
the local intelligent inference on mobile devices like smartphones. Recent
study~\cite{xu2019first,sun2021mind,almeida2021smart} indicates the \textit{intelligent Apps} (iApps),
smartphone Apps using in-App DL models, will be increasingly popular, which is also verified by our own study shown
in Section~\ref{subsec:iApp_statis}.

\begin{figure}[t]
    \setlength{\abovecaptionskip}{2pt}
	  \includegraphics[width=0.35\textwidth]{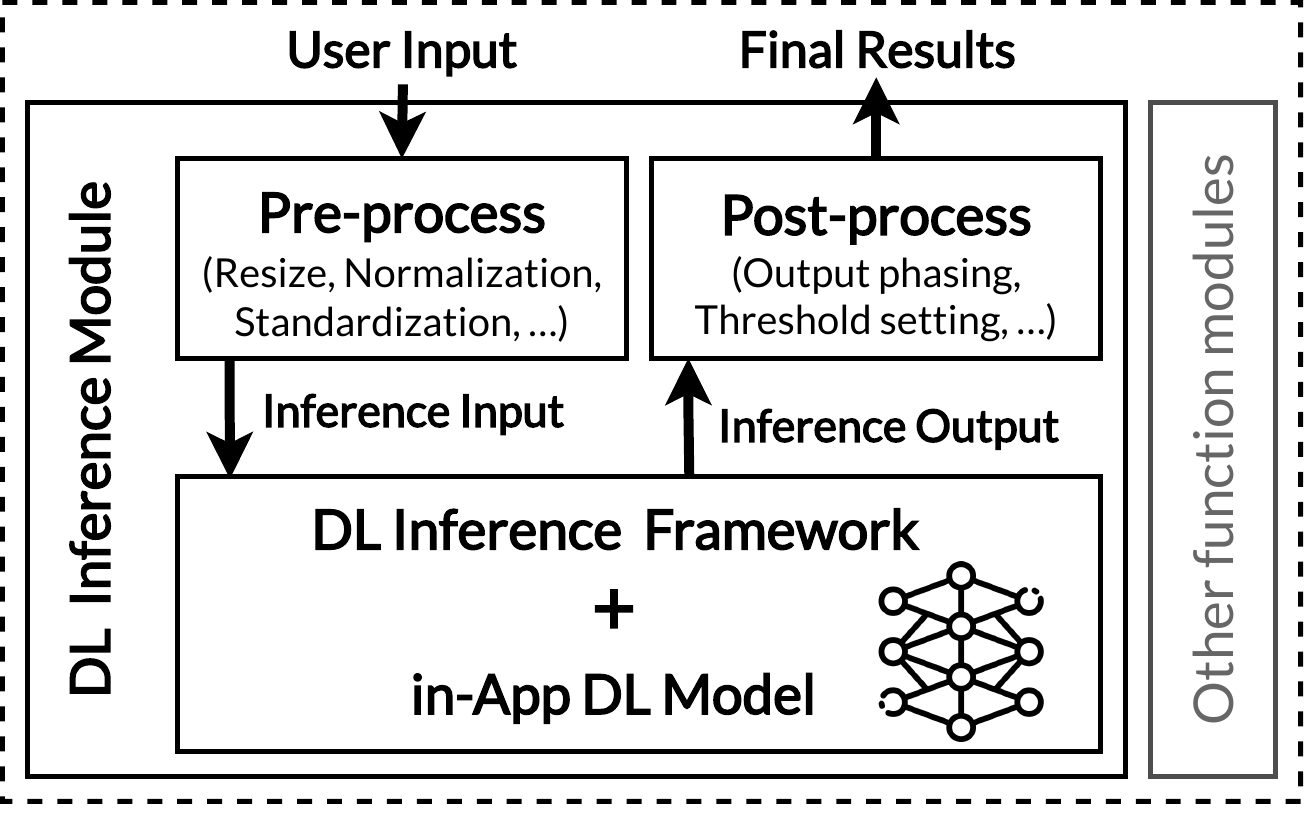}
  \caption{A typical structure of an iApp. Compared to ordinary Apps, the key difference is a software component
  performing the local intelligent inference. This component consists of a in-App DL model optimized for inference and
  its paired code for input and output processing. App marketplaces so far do not support the auto testing of this
  component from the AI security perspective.}
  \label{fig:protection_scenario}
  \vspace{-10pt}
\end{figure}

The key difference between an iApp and an ordinary App is a software component performing the local intelligent inference
(Figure~\ref{fig:protection_scenario}). This component usually consists of two parts, the \textit{in-App DL model} and
\textit{IO processing code}. The in-App DL model is commonly optimized for easy deployment and speedy inference, and
thus gets rid of the backpropagation (BP) ability. The IO processing code is tightly associated with the corresponding in-App
model, and it is responsible for both preparing inference inputs and interpreting inference outputs. Incorrect IO
processing will impede the success of DL inference.

Such key difference in iApps brings troubles to the App marketplaces' important routine work --- the
security-oriented auto testing of Apps released to them. This is because, by nature, the testing philosophy,
paradigm, and requirements for neural networks are different from those for software
codes~\cite{gao2020backdoor,huang2020survey,wang2019neural}. Regarding an in-App DL model, the inference changes to
perturbed inputs, including their possible interpretations, are more important than the behavior changes of codes. For
example, a road-lane-detection iApp is insecure and should be delisted from App marketplaces if it is too easy to show
different lane lines when adversarial perturbations, say a small mark on the road, are applied. Since more and more Apps
go intelligence, it is urgent to resolve how to efficiently auto-test massive released iApps in a comprehensive manner
from the AI security perspective. However, to remove such "dark cloud" of auto-testing requires a new tool design
addressing two challenges.

\textit{Model Conversion}. BP operations on the DL model are not only inevitable in various model testing methods~\cite{madry2017towards,
kurakin2016adversarial, dong2018boosting, chen2018ead}, but also critical to the DL interpretability~\cite{zhang2018visual,li2021interpretable}.
However, it is difficult to convert an in-App DL model into its white-box counterpart on a BP-enabled test platform, because
such a model in App is heavily optimized for inference. For example, the neural operators' attributes are implicit
and hidden in the App codes, and weights quantization makes gradient computations extremely complicated. 

\textit{Reliable Slicing}. A converted white-box model cannot properly run on the testing platform without the cooperation of its own IO processing
code in the App. Therefore, the App marketplace needs to precisely slice out this part of the released App code and re-use it during the
testing. However, this code slicing should be efficient and accurate with an extremely high success rate, so that the massive testing of
released iApps are able to be supported with no manual efforts involved.
For example, IO processing often needs to perform assignment operations between data objects and arrays, which is hard to be tracked accurately in general cases.




Existing testing tools cannot meet the above needs. First, existing model conversion tools, e.g.,
MMdnn~\cite{liu2020enhancing} and ONNX-based~\cite{onnx} tools, can translate trainable DL models between frameworks,
but they cannot translate inference-only models into corresponding BP-enabled versions. Second, the dynamic code
slicing~\cite{azim2019dynamic} needs human involvement, while the success rate of static code slicing is not high. 
We find that the Android slicing tool Jicer~\cite{pauck2021jicer} cannot extract runnable IO processing codes for model testing purpose.

In this work, we propose an automated tool, \sysname, which can enable large-scale \textbf{\underline{T}}esting of
\textbf{\underline{I}}nApp DL \textbf{\underline{M}}odels for App marketplaces. \sysname can successfully extract the
in-App DL models from massive real-world iApps and its IO processing code, and then prepares the ready-to-test versions of in-App DL models with IO procesing code for various model assessments.
The whole procedure of \sysname is efficient without developers' cooperation like model information or source code. The
outputs are capable of BP operations which are important to many model assessments and interpretation methods. \sysname
can help App marketplaces to auto-test iApps in practice from the AI security perspective.

The key designs of \sysname are two reconstruction techniques, i.e., precise code reconstruction and BP-enabled model reconstruction. The code reconstruction adopts a precise Android static slicing, which can produce runnable processing codes by fully considering the App's control flow, invoking context, field access, and branch selection during the slicing. It also proposes an IO-processing-oriented code generation to ensure that sliced statements can execute in the same order as they are in the original iApp.
The BP-enabled model reconstruction can convert an inference-only model into a white-box one, which is used for comprehensive testing. It first abstracts the framework-independent computation procedure and then utilizes a rule-based operator's attributes completion of recovering the stripped information for model reconstruction.

We implement the \sysname and test in-App models at scale. We collect about 15k unique Apps from five marketplaces. Following the in-App model finding method proposed by the work~\cite{xu2019first}, we find there are 3,064 iApps equipped with 800 unique in-App DL models. Then we test 100 unique in-App models with popular DL frameworks in terms of robustness. We find 56\% of the in-App models are vulnerable to robustness issues (Section~\ref{subsec:ra}). \sysname also detects physical attacks against three representative in-App models. Details of the measurement are presented in Section~\ref{sec:oma}.

We highlight the key contributions as follows:

\begin{enumerate}
	\item \sysname is the first effort to enable in-App DL model testing at scale. It can automatically reconstruct the IO processing code and BP-enabled DL models for powerful white-box testing techniques. \sysname is fully-automatic and works without the iApp provider's cooperation.
	\item \sysname proposes two novel reconstruction techniques. The code reconstruction is a precise Android static slicing to produce runnable IO processing code. The model reconstruction rebuilds the BP-enabled model from its inference-only version by establishing equivalent calculation and information completion.
	\item  We perform robustness assessment on 100 \textit{unique commercial in-App models} through \sysname. We also successfully detect physical adversarial attacks against commercial iApps, which may cause economic losses and serious security issues.
\end{enumerate}

%% file: background.tex
\section{Background and Related Works}
\label{sec:bg}

\subsection{In-App DL Models}
\label{subsec:iApps}

The iApps equipped with DL models are emerging. The in-App models have been studied preliminarily. The work~\cite{xu2019first} is the first work to keep an eye on iApps and proposes a way to find DL models and frameworks in an App. Then the work~\cite{sun2021mind} investigates how iApp providers protect in-App DL models. The work~\cite{almeida2021smart} and work~\cite{zhang2022comprehensive} perform comprehensive studies on DL model inference performance. The work~\cite{huang2021robustness} studies the robustness of the in-App DL models. 

Existing works do not propose techniques for automated reconstruction of IO processing code in the iApp. These works either do not need the IO processing code or reconstruct the processing code manually~\cite{xu2019first,almeida2021smart}. These works can also not directly reconstruct the BP-enabled model from the in-App model to perform the model assessment. For example, the work~\cite{huang2021robustness} assesses the model's robustness by utilizing the adversarial attacks' transferability.

\subsection{Android Slicing}
\label{subsec:as}

Programing slicing is a technique to extract statements that may affect a given statement (\texttt{stmt}) and a set of values (\texttt{V}) through data dependency and control dependency analysis. The given statement and the set of values are referred as a \textit{slicing criterion} <\texttt{stmt,V}>. There are two well-known frameworks in the Android analysis scenarios, i.e., WALA~\cite{sridharan2007thin} and Soot~\cite{lam2011soot}. The frameworks support the basic functionalities and programable interfaces to build slicing algorithms. The code reconstruction of \sysname is based on Soot because Soot has better support for Dalvik bytecode, and many Androids analysis tools are developed based on Soot. 

The work~\cite{azim2019dynamic} designs a human-involved Android dynamic slicing tool by which users first perform App instrumentation, manually run the instrumented App, and dynamically collect logs to perform Android slicing. AppSlicer~\cite{bhardwaj2019serving} is another activity-level App dynamic slicing technique. It first triggers all activities in a simulator, records the code invoked by each activity, and slices the target activity according to the recorded content. Jicer~\cite{pauck2021jicer} is a work to perform general Android static slicing. It aims at slicing the code of any App with any slicing criterion. However, Jicer fails to slice the processing code from a commercial iApp because it lacks the processing code oriented design. 

The slicer's accuracy depends on its sensitivity regarding various program features~\cite{pauck2021jicer}. A flow-sensitive slicer takes control dependencies into full consideration. Fully considering from which context a method is called makes the slicer context-sensitive. Handling data dependencies between usages of the same field in different methods makes a slicer field-sensitive. When handling conditional statements, a path-sensitive slicer must consider slicing which branch. 

The code reconstruction of \sysname is a fully-automatic precise Android static slicing without human involvement. In addition, our reconstruction tool is \textit{flow-}, \textit{context-}, \textit{field-}, \textit{path-}sensitive.

\subsection{Adversarial Attacks}
\label{subsec:ae}

The adversarial attacks have been widely studied in both the AI and security communities. DL models have been demonstrated that they are vulnerable to adversarial examples~\cite{szegedy2013intriguing}. The adversaial examples are inputs to DL models that have been intentionally optimized to cause models to make a mistake. Specifically, given a DL model $f_{\theta}(\cdot)$ with parameters $\theta$ and an input $x$ with a ground truth label $y$, an adversarial exmaples $x'$ is produced by optimization, which is closed to $x$. $x'$ is able to cause the DL model to make an incorrect prediction as $f_{\theta}(x') \neq y$ (\textit{untargeted attacks}), or $f_{\theta}(x') = y*$ (\textit{targeted attacks}) for some $y* \neq y$. 

The attack scenarios can be classified by the amount of knowledge the adversary has about the model, i.e., \textit{white box attack}~\cite{madry2017towards} and \textit{black box attack}~\cite{brendel2017decision}. In the white box scenario, the adversary fully knows the model, including model type, model architecture, and values of all parameters. The adversary can perform gradient-based attacks on the model. In the black box scenario, the adversary has limited information about the model. The adversaries can only perform the attack by probing and observing the output.

The adversarial attacks can be classified into \textit{digital attacks}~\cite{goodfellow2014explaining} and \textit{physical attacks}~\cite{eykholt2018robust} according to how the adversary modify the input data. In a digital attack, the adversary has direct access to the actual data fed into the model. The adversary can modify each bit of the input data. In a physical attack, the adversary does not have direct access to the digital representation of the input data. The adversary can place objects in the physical environment seen by the camera. 

In our work, we assess the robustness of the in-App models through the white box digital attacks. We also perform white box physical attacks on three representative in-App models.

%% file: overview.tex
\section{Overview}
\label{sec:sys_design}

In this section, we first define the security model and design requirements of \sysname. Then we present the high-level design of \sysname and introduce the two key reconstruction techniques.

\subsection{Problem Overview}
\label{subsec:probOverview}

The security issues of in-App models are able to cause the iApp to behave abnormally. And the in-App model may even become a "protective umbrella" for the iApp's malicious behavior. Therefore, App markets have a strong need to test the in-App models in massive iApps.

\subsubsection{Security Model.}
The iApp may be developed by careless or malicious developers. Those careless developers may use an in-App model that has not undergone comprehensive testing. Those malicious developers even launch active attacks by poisoning the training data. We do not directly handle the scenario where the iApp is packed. If App markets want to test the in-App model in a packed iApp, they can first use previous works~\cite{duan2018things,xue2020packergrind,xue2021happer} to unpack the iApp, and then use \sysname to prepare the IO processing code and BP-enabled model.

\subsubsection{Requirements.}
\label{subsubsec:reqs}
Given the challenges discussed in Section~\ref{sec:intro}, 
we summarize three requirements to perform comprehensive testing on in-App models.


\textit{\textbf{First}, the model before and after reconstruction should be equivalent.} The model reconstruction mainly performs two things. One is porting the in-App model to a DL framework that supports training. The other is removing the obstacles that hinder gradient computation, such as undoing weights quantization and replacing the inference-only operations. The reconstructed model and the in-App model should be perfectly equivalent in computational procedure and trainable parameters so that testing on the reconstructed model is equivalent to testing on the in-App model.

\begin{figure}[t]
\setlength{\abovecaptionskip}{2pt}
  \includegraphics[width=0.45\textwidth]{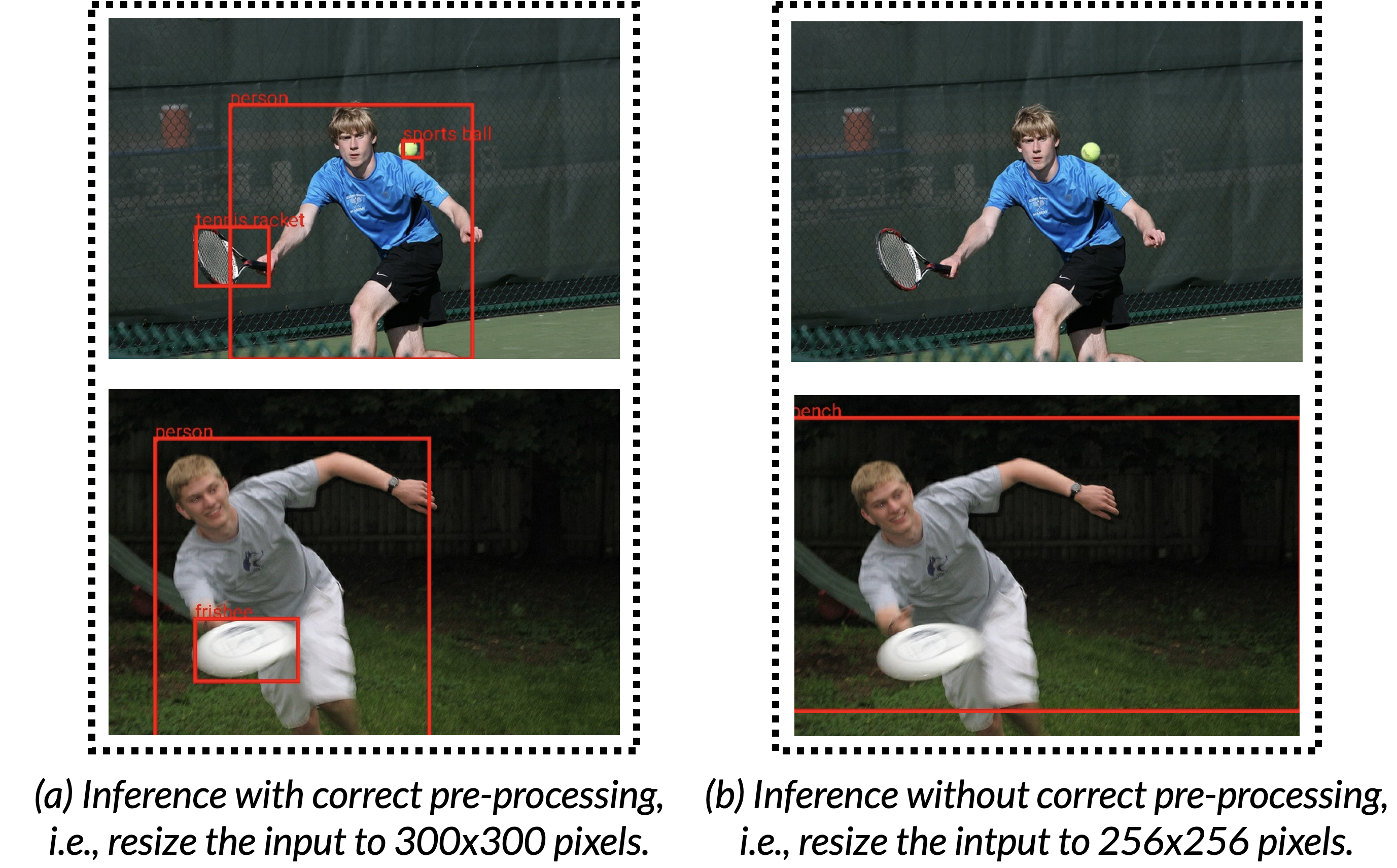}
  \caption{The inference results on the user data with and without correct pre-processing. \label{fig:pre-process_case}}
  \vspace{-10pt}
\end{figure}

\textit{\textbf{Second}, the processing code reconstruction should be precise and runnable.} 
We show a case\footnote{The DL model is taken from an iApp, whose bundle ID is org.prudhvianddheeraj.lite.example.detection.} in Figure~\ref{fig:pre-process_case} to demonstrate the effect of processing code on the correctness of DL inference. 
Before being fed into the DL model, the user input needs to be resized in a preset way. If the resize operation is omitted or substituted by a random resize operation, the inference result is wrong. The reconstruction should be precise so that processing operations are not missed. At the same time, the reconstruction should be practical so as to ensure that it can produce runnable results.

\textit{\textbf{Thrid}, the proposed tool should be automatic.}
To enable App markets to test in-App models at scale, the model and code reconstructions should be fully automatic without human involvement. And in our scenario, the reconstructions should be done without the cooperation of iApp developers.

\begin{figure*}[t]
\setlength{\abovecaptionskip}{2pt}
  \includegraphics[width=\textwidth]{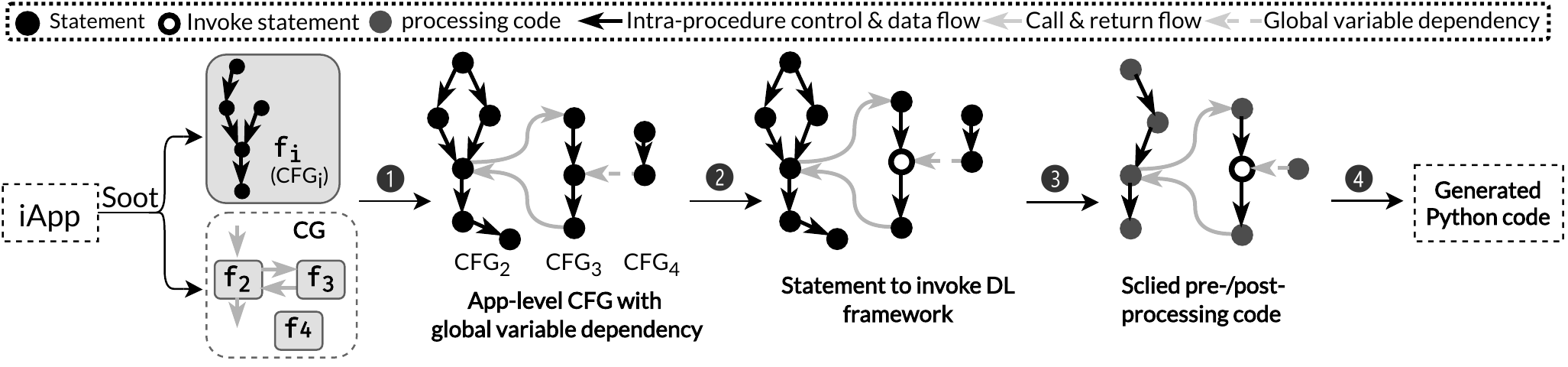}
  \caption{The workflow of the code reconstruction. \label{fig:cr} }
  \vspace{-10pt}
\end{figure*}

\subsection{\sysname Design Overview}
\label{subsec:des_ow}

In this section, we introduce the \sysname's workflow and explain how \sysname meets the requirements discussed in Section~\ref{subsec:probOverview} in brief. 

We first used the work~\cite{xu2019first} to determine whether an App contains in-App models. If an iApp is found, we then perform the proposed code reconstruction and model reconstruction to build the test object. 

\textit{During the code reconstruction}, \sysname takes as input a released iApp and outputs runnable IO processing codes. It first finds the invoking statement of the DL inference framework through static program analysis. Note that we do not need to reconstruct the DL inference framework because it will be replaced by a ready-made framework that supports BP.  

Then, \sysname performs precise Android static slicing on the released iApp to extract the IO processing code. The slicing starts at the DL framework's invoking statement, and it performs backward slicing to extract all statements that determine the value used by the invoking statements and performs forward slicing to extract all statements that use the value defined by the invoking statements. 

Next, \sysname generates Python code that can run on PCs based on the sliced IO processing code for model testing. Generating Python code from the sliced APK is feasible because the sliced codes are responsible for IO processing, which is not coupled with the Android platform and system services. 

\textit{During the model reconstruction}, \sysname takes as input an in-App model and outputs its BP-enabled version. \sysname first extracts the computation procedure of the in-App model and eliminates the factors hindering the gradient calculation by undoing the quantization, removing inference-specific operations, and so on. 

Then, \sysname utilizes a set of carefully elaborate rules to rebuild the stripped information during the deployment-oriented conversion. The stripped information is mainly the attributes of the operations, such as the filter numbers and kernel size of a convolution operation. Finally, with the rebuilt computation procedure and stripped attributes, \sysname reconstructs the BP-enabled model with a widely-used DL framework Keras\footnote{\url{https://keras.io/}}. 

Combining the code reconstruction and model reconstruction, our \sysname can produce a testing object, equivalent to the iApp's DL inference module, that can be assessed by various powerful testing techniques. In the next section, we will detail the \sysname's design. And in Section~\ref{sec:oma}, we will show how to use \sysname to test the in-App models of commercial iApps at scale.

%% file: code_recon.tex
\section{Precise Code Reconstruction}
\label{sec:cr}

To reconstruct the IO processing code, \sysname performs precise Android static slicing on the iApp. It is able to extract all user input and inference output processing code from the iApp and generate the corresponding Python code for the following testing. The pre-processing code can be extracted by iteratively finding all statements that determine the parameters of the DL framework inference interface. The post-processing code can be extracted by finding all statements tainted by the inference output. 

To achieve the above objectives, \sysname performs both backward and forward slicing starting from the DL framework invoking statements to find all IO processing codes. The backward slicing stops once semantically explicit user input is found, such as an unprocessed image or sound. The forward slicing stops once the inference output is parsed into a structure that can be used for model testing, such as category and confidence.


We demonstrate the code reconstruction workflow in Figure~\ref{fig:cr}. The following sections detail the design of the code reconstruction. 

\subsection{Slicing Preparation}
\label{sbsec:datapre}

\begin{figure}[t]
  \includegraphics[width=0.42\textwidth]{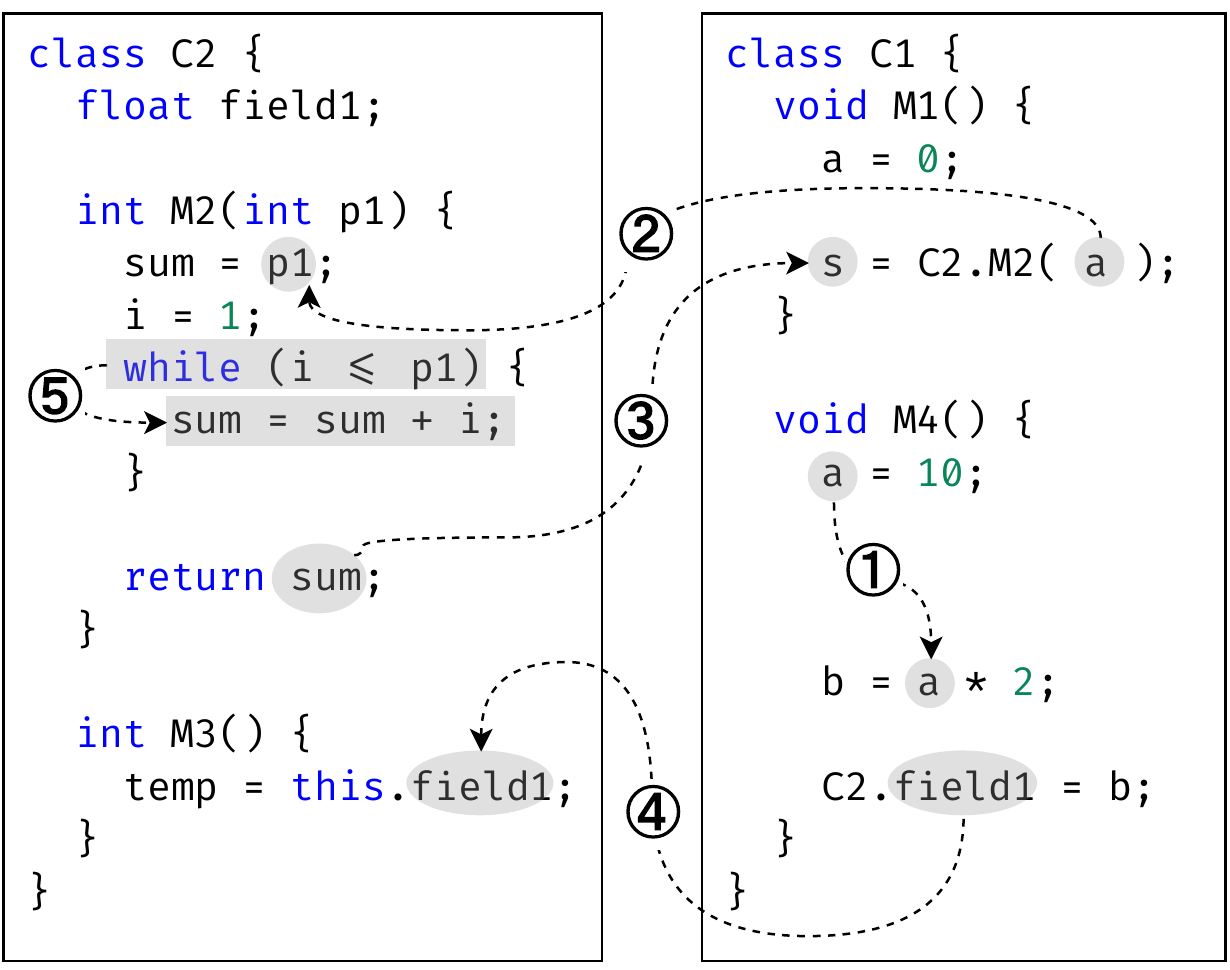}
  \caption{Control and data dependency among the statements considered by \sysname. \label{fig:ex_pdg}}
  \vspace{-10pt}
\end{figure}

After taking as input, an iApp, \sysname utilizes Soot to build the App's call graph (CG) and function-level control flow graphs (CFG). 
\sysname builds an App-level CFG with global variable dependency for the given iApp (Step\ding{182} in Figure~\ref{fig:cr}).  In this step, \sysname adds edges that \textit{1)} represent dependency between the caller and callee according to the CG (e.g., Edge~\ding{193} and Edge~\ding{194} in Figure~\ref{fig:ex_pdg}), and \textit{2)} represents dependency between the statements that read or write the same field in an class (e.g., Edge~\ding{195} in Figure~\ref{fig:ex_pdg}). 

We denote the built graph as the \textit{slicing basis} because it is able to represent data and control dependencies between statements in an iApp. The data dependency and control dependency between two statements are denoted as $\leftarrow_{d}$ and $\leftarrow_{c}$, respectively. 

As for the data dependency, if \texttt{use(stmt$_1$)} $\cap$ \texttt{def(stmt$_2$)} $\neq$ $\emptyset$, \texttt{stmt}$_1$ $\leftarrow_d$ \texttt{stmt}$_2$. The data dependencies we consider can be classified into two types. The first type of data dependency is brought by the def-use relationship of local variables in a single function (Edge\ding{192} in Figure~\ref{fig:ex_pdg}). The second type of data dependency is brought by the def-use relationship of filed variables between functions (Edge\ding{195} in Figure~\ref{fig:ex_pdg}). 

We consider three types of control denpendency. The control dependency in a single function is brought by the procedure branching statement. If \texttt{stmt$_1$} can determine whether \texttt{stmt$_2$} is executed, \texttt{stmt}$_2$ $\leftarrow_c$ \texttt{stmt}$_1$ (Edge\ding{196} in Figure~\ref{fig:ex_pdg}). The control dependency between functions occurs when a function is called (Edge\ding{193} in Figure~\ref{fig:ex_pdg}) or exits (Edge\ding{194} in Figure~\ref{fig:ex_pdg}).

After building the slicing basis, \sysname prepares the slicing criterions for the following App slicing (Step\ding{183} in Figure~\ref{fig:cr}). Due to different inference tasks and different engineering implementations, the entry point and the exit point of IO processing lack uniform characteristics among different iApps. Therefore, it is difficult for us to locate the entry point and exit point of the IO processing as the slicing criterions. In contrast, the DL framework's invoking interfaces, which are used to load DL models and perform universal DL computation, have significant static characteristics~\cite{zhang2022comprehensive}. \sysname utilizes these invoking interfaces to prepare slicing criterions for the forward and backward slicing, respectively. The backward slicing criterions consist of statements that invoke the DL framework and the values used by the corresponding statement. The goal of backward slicing is to extract all statements that determine the used values. The forward slicing criterions consist of statements that invoke the DL framework and the values defined by the corresponding statement. The goal of forward slicing is to extract all statements that use the defined values.

To find the invoking interfaces of DL framework, we improve the idea used by existing works~\cite{xu2019first,sun2021mind}. If the iApp utilizes an open-sourced DL framework, e.g., TFLite, with well-documented interfaces, we can directly locate statements that invoke the interfaces in slicing basis as the slicing criterions. We find using the invoking interfaces' parameters and return values and the availability of the invoking methods is robust enough to determine the invoking interfaces when the iApp is protected with string-based obfuscation.

\vspace{-10pt}
\subsection{Code Slicing}

\sysname proposes an Android slicing technique to extract the user data pre-processing and inference output post-processing code (Step\ding{184} in Figure~\ref{fig:cr}). The slicing technique can perform bidirectional slicing starting from the found slicing criterions. 

We denote the slicing criterion as \texttt{<stmt,V>}. The \texttt{stmt} in the slicing criterion denotes the invoking statement of DL framework, which is the slicing start point. \texttt{V} represents variables used by \texttt{stmt} when preforming the backwards slicing, i.e., \texttt{V} = \texttt{use}(\texttt{stmt}). \texttt{V} represents variables defined by \texttt{stmt} when preforming the forwards slicing, i.e., \texttt{V} = \texttt{def}(\texttt{stmt}). For example, \texttt{int[] r = function(String p1, byte[] p2, long[] p3)} is a DL framework invoking statement, denoted as \texttt{stmt}$_\texttt{criterion}$. Its def values and use values are \texttt{\{r\}} and \texttt{\{p1, p2, p3\}}.

The slicing algorithm is shown in Algorithm~\ref{alg:slicing}. The algorithm takes as input the slicing basis and extracts statements belonging to the processing code by analyzing the data and control dependency. Recall that the slicing basis is the App-level CFG with global variable dependency built in Section~\ref{sbsec:datapre}. Each edge of the slicing basis graph represents the data or control dependency between two statements.

When performing backward slicing, if the statement to analyze has a data or control dependency with the sliced statement (line 11), the statement will be added to the sliced results (line 12). And the values defined by the newly-sliced statement can be removed from the slicing criterion, and the values used by the newly-sliced statement should be added to the slicing criterion (line 13). Then the backward slicing goes on (line 14). 

When performing forward slicing, if the statement to analyze does not have a data or control dependency with the sliced statement (line 21$\sim$22), the values defined by the statement will be removed from the slicing criterion (line 23). 
If the statement to analyze has a dependency on the sliced statement, the value defined by the statement should be added to the slicing criterion (line 24). In addition to continuing the forward slicing, we should also perform the backward slicing to determine the values used by the newly-added statement (line 29$\sim$30, 32).

The slicing ends when all the values can be determined (line 2$\sim$3). Or there are no more statements to analyze (line 7$\sim$8). 
The $pred\_of(\cdot)$ (line 6) and $succ\_of(\cdot)$ (line 16) represent the predecessor and successor of the given statement on the slicing basis.

\begin{algorithm}[t]
\DontPrintSemicolon
\small
\KwData{The App-level CFG with global variable dependency, $\texttt{G}_b$; The slicing criterion, <\texttt{stmt}$_{sc}$,$V_{sc}$>; The slicing direction, \texttt{D}.}
\KwResult{The slicing results: $\texttt{S}_{res}$.}
\Begin{
    \If{ V$_{sc}$ == $\emptyset$} {
        \KwRet{$\texttt{S}_{res}$}\;
    }
    ~\;
	\If{ \texttt{D} == "backward" }{
	    $\texttt{stmt}_n = \texttt{G}_b.pred\_of(\texttt{stmt}_{sc})$\;
	    \If{ $\texttt{stmt}_n$ == NULL} {
            \KwRet{$\texttt{S}_{res}$}\;
        }
        $V_d = def(\texttt{stmt}_n)$ \;
        $V_u = use(\texttt{stmt}_n)$ \;
        \If{$V_d \cap V_{sc} \neq \emptyset$}{
            $\texttt{S}_{res}.add(\texttt{stmt}_n)$\;
            $V_{sc}' = (V_{sc} - V_d) \cup V_u$\;
        } 
        $\texttt{S}_{res}.add$(\texttt{slice}($\texttt{G}_b$, <\texttt{stmt}$_n$,$V_{sc}'$>, "backward"))\;
    }

   	\If{ \texttt{D} == "forward" }{
	    $\texttt{stmt}_n = \texttt{G}_b.succ\_of(\texttt{stmt}_{sc})$\;
	    \If{ $\texttt{stmt}_n$ == NULL} {
            \KwRet{$\texttt{S}_{res}$}\;
        }
        $V_d = def(\texttt{stmt}_n)$ \;
        $V_u = use(\texttt{stmt}_n)$ \;
        \If{$V_u \cap V_{sc} == \emptyset$}{
            \If{$V_d \cap V_{sc} \neq \emptyset$}{
                $V_{sc}' = V_{sc} - V_d$\;
            }
            $\texttt{S}_{res}.add$(\texttt{slice}($\texttt{G}_b$, <\texttt{stmt}$_n$,$V_{sc}'$>, "forward"))\;
        }
        \Else {
            $\texttt{S}_{res}.add(\texttt{stmt}_n)$\;
            $V_{sc}' = V_{sc} \cup V_d$\;
            $\texttt{S}_{res}.add$(\texttt{slice}($\texttt{G}_b$, <\texttt{stmt}$_n$,$V_{sc}'$>, "forward"))\;
            
            $V_{sup} = V_u-V_{sc}$\;
            \If{ $\texttt{stmt}_p$ == NULL} {
                \KwRet{$\texttt{S}_{res}$}\;
            }
            $\texttt{S}_{res}.add$(\texttt{slice}($\texttt{G}_b$, <\texttt{stmt}$_{n}$,$V_{sup}$>, "backward"))\;
        }
    }
}
\caption{Pseudo code of performing slicing, denote as \texttt{slice($\cdot$)}.}\label{alg:slicing}

\end{algorithm}

\subsection{Code Generation}

After slicing all statements, \sysname generates the executable python code for the following DL model assessment (Step\ding{185} in Figure~\ref{fig:cr}). The code generation is feasible because the sliced code is for data processing in the DL inference scenario, which is not Android platform-specific. The generation procedure consists of two steps, i.e., statement translation and statement ordering.


\subsubsection{Statement Translation.}

The code slicing works on an intermediate representation (IR) provided by Soot \footnote{\sysname utilizes Jimple, which is the widely-used intermediate language in Android App analysis.}. Most of the IR code, e.g., assignment statement and invoking statement, can be easily translated into python code. The IR code needing elaborate processing is the conditional statement, i.e., if statement, and jump statement, i.e., goto statement.
This is because the loop structure is also represented as the combination and nesting of the conditional statement and jump statement. We should build the loop structure, especially the loop condition and loop body, to ensure the correctness of data processing. 

\sysname adopts the following rule to identify the loop structure. We denote a sliced statement as \texttt{stmt} and the CFG of the function that \texttt{stmt} belongs to as \texttt{G}. The \texttt{stmt}'s direct successor statement is \texttt{stmt}$_{succ}$ on the \texttt{G}. If on the \texttt{G}, \texttt{stmt}$_{succ}$ dominates \texttt{stmt}\footnote{Note that, in an CFG, a statment $\texttt{stmt}_1$ dominates a statment $\texttt{stmt}_2$, if every path from the CFG's entry statement to $\texttt{stmt}_2$ must go through $\texttt{stmt}_1$.}, we say a loop structure is found. and the \texttt{stmt}$_{succ}$ is the entry point of the loop body. The loop body of the found loop structure is the intersection of all predecessor statements of \texttt{stmt} and all successor statments of \texttt{stmt}$_{succ}$. The loop condition of the found loop structure is the if statment in the loop body whose target statement\footnote{The \textbf{\texttt{stmt'}} is the target statement of an if statement \textbf{\texttt{if condition goto stmt'}}.} is beyond the loop body. Then we can reconstruct all loop structures in sliced code and then translate them into Python.

\subsubsection{Statement Ordering}

The order in which the sliced statements executes determines the correctness of the data processing. We first organize the sliced statements in terms of functions as they are in the original iApp. The order of the statements can be determined according to the CFG of the corresponding function. Next, we define all the organized functions and then determine how to call these functions in the correct order.

According to \sysname's design, these organized functions have explicit invoking relationships or implicit data dependencies with at least one of the rest functions. Note that the data dependency is brought about by reading or writing the same field. We group these functions according to whether there is a direct or indirect invoking relationship between them. For example, if $f_A()$-$call$->$f_B()$-$call$->$f_C()$, then $f_A()$, $f_B()$, and $f_C()$ are organized into one function group. We denote the function without any caller in each function group as head functions. In the above example, $f_A()$ is a head function. Now, we only need to order these head functions, and the rest functions in the function groups will automatically be invoked when the head function is called.

To order the head functions, we propose a "Write-before-Read" principle by fully considering the field dependencies among the functions. The principle consists of two rules.
\begin{enumerate}
	\item[\textbf{R1:}] Function groups that do not read any field can be arranged in any order.
	\item[\textbf{R2:}] For any field, the function group that writes a field should rank before the function group that reads that field.
\end{enumerate}

\sysname first records the filed variable reads and writes of each function group, denoted as $gv_f^R$ and $gv_f^W$, respectively. 
The subscript $f$ indicates the function group. Then we arrange the head functions whose function group's $gv_f^R$ is empty in arbitrary order (\textbf{R1}). Once a head function is organized, \sysname updates $gv_g^R$ of each of the rest function groups by $gv_g^R= gv_g^R - gv_f^W$. The \sysname will repeatedly order function groups whose $gv^R$ is empty (\textbf{R2}) until all function groups are arranged.

%% file: model_recon.tex
\section{BP-enabled Model Reconstruction}
\label{sec:mr}

\begin{table*}[]
\small
\begin{tabular}{|c|c|c|c|}
\hline
\textbf{Operator Type}                                                                  & \textbf{Attribute} & \textbf{Needed Information}                                                             & \textbf{Computation Rule}                                                                                                                                                                                                                                                                     \\ \hline
\multirow{6}{*}{\begin{tabular}[c]{@{}c@{}}Conv2D\\ {[}DepthwiseConv2D{]}\end{tabular}} & filters            & \texttt{output\_shape}                                                                               & \texttt{output\_shape}{[}-1{]}                                                                                                                                                                                                                                                                            \\ \cline{2-4} 
                                                                                        & kernel\_size{[}0{]} & \texttt{weight\_shape}                                                                               & \texttt{weight\_shape}{[}1{]}                                                                                                                                                                                                                                                                             \\ \cline{2-4} 
                                                                                        & kernel\_size{[}1{]} & \texttt{weight\_shape}                                                                               & \texttt{weight\_shape}{[}2{]}                                                                                                                                                                                                                                                                             \\ \cline{2-4} 
                                                                                        & strides{[}0{]}     & \begin{tabular}[c]{@{}c@{}}\texttt{input\_shape}\\ \texttt{output\_shape}\\ kernel\_size\end{tabular}           & $round(\frac{\text{\texttt{input\_shape}[1]} - \text{kernel\_size[1]}}{\text{\texttt{output\_shape}[1]} - 1})$                                                                                                                                                                                                       \\ \cline{2-4} 
                                                                                        & strides{[}1{]}     & \begin{tabular}[c]{@{}c@{}}\texttt{input\_shape}\\ \texttt{output\_shape}\\ kernel\_size\end{tabular}           & $round(\frac{\text{\texttt{input\_shape}[2]} - \text{kernel\_size[2]}}{\text{\texttt{output\_shape}[2]} - 1})$                                                                                                                                                                                                       \\ \cline{2-4} 
                                                                                        & padding            & \begin{tabular}[c]{@{}c@{}}\texttt{input\_shape}\\ \texttt{output\_shape}\\ kernel\_size\\ strides\end{tabular} & \begin{tabular}[c]{@{}c@{}}"same" when \\ $\lfloor \frac{\text{\texttt{input\_shape}[i]} - \text{kernel\_size[i-1]}}{\text{strides[i-1]}} + 1 \rfloor = \text{\texttt{input\_shape}[i]} (i=1,2)$ \\ {[}default: "valid"{]}\end{tabular}                                                                              \\ \hline
DepthwiseConv2D                                                                         & depth multiplier   & \begin{tabular}[c]{@{}c@{}}\texttt{input\_shape}\\ \texttt{output\_shape}\end{tabular}                         & \texttt{output\_shape}{[}-1{]} / \texttt{input\_shape}{[}-1{]}                                                                                                                                                                                                                                                      \\ \hline
\multirow{6}{*}{Conv2DTranspose}                                                        & filters            & \texttt{output\_shape}                                                                               & \texttt{output\_shape}{[}-1{]}                                                                                                                                                                                                                                                                            \\ \cline{2-4} 
                                                                                        & kernel\_size{[}0{]} & \texttt{weight\_shape}                                                                               & \texttt{weight\_shape}{[}1{]}                                                                                                                                                                                                                                                                             \\ \cline{2-4} 
                                                                                        & kernel\_size{[}1{]} & \texttt{weight\_shape}                                                                               & \texttt{weight\_shape}{[}2{]}                                                                                                                                                                                                                                                                             \\ \cline{2-4} 
                                                                                        & strides{[}0{]}     & \begin{tabular}[c]{@{}c@{}}\texttt{input\_shape}\\ \texttt{output\_shape}\\ kernel\_size\end{tabular}           & $round(\frac{\text{\texttt{output\_shape}[1]} - \text{kernel\_size[1]}}{\text{\texttt{input\_shape}[1]} - 1})$                                                                                                                                                                                                       \\ \cline{2-4} 
                                                                                        & strides{[}1{]}     & \begin{tabular}[c]{@{}c@{}}\texttt{input\_shape}\\ \texttt{output\_shape}\\ kernel\_size\end{tabular}           & $round(\frac{\text{\texttt{output\_shape}[2]} - \text{kernel\_size[2]}}{\text{\texttt{input\_shape}[2]} - 1})$                                                                                                                                                                                                       \\ \cline{2-4} 
                                                                                        & padding            & \begin{tabular}[c]{@{}c@{}}\texttt{input\_shape}\\ \texttt{output\_shape}\\ kernel\_size\\ strides\end{tabular} & \begin{tabular}[c]{@{}c@{}}"same" when \\ $ \lfloor (\text{\texttt{input\_shape}[i]},- 1)*\text{strides[i-1]} + \text{kernel\_size[i-1]} \rfloor =$ \\ $\text{\texttt{output\_shape}[i]}$ (i=1,2) {[}default: "valid"{]}\end{tabular}                                                                                   \\ \hline
\multirow{2}{*}{\begin{tabular}[c]{@{}c@{}}MaxPooling/\\ AveragePooling\end{tabular}}   & pool\_size          & \begin{tabular}[c]{@{}c@{}}\texttt{input\_shape}\\ \texttt{output\_shape}\end{tabular}                         & $round(\frac{\text{\texttt{input\_shape}[1]}}{\text{\texttt{output\_shape}[1]}})$                                                                                                                                                                                                                                   \\ \cline{2-4} 
                                                                                        & padding            & \begin{tabular}[c]{@{}c@{}}\texttt{input\_shape}\\ \texttt{output\_shape}\\ pool\_size\end{tabular}             & \begin{tabular}[c]{@{}c@{}}"same" when \\ $\lfloor \frac{\text{\texttt{input\_shape}[i]} }{\text{pool\_size} + \varepsilon}  \rfloor < \text{\texttt{output\_shape}[i]}$\\ (i=1,2) {[}default: "valid"{]}\end{tabular}                                                                                               \\ \hline
UpSampling                                                                              & size               & \begin{tabular}[c]{@{}c@{}}\texttt{input\_shape}\\ \texttt{output\_shape}\end{tabular}                         & $\lfloor \frac{\text{\texttt{input\_shape}[1]}}{\text{\texttt{output\_shape}[1]}} \rfloor$                                                                                                                                                                                                                          \\ \hline
Pad, MirrorPad                                                                          & padding            & \begin{tabular}[c]{@{}c@{}}\texttt{input\_shape}\\ \texttt{output\_shape}\end{tabular}                         & \begin{tabular}[c]{@{}c@{}}$\text{padding[i][0]} = \lfloor \frac{\text{\texttt{output\_shape}[i]} - \text{\texttt{input\_shape}[i]}}{2} \rfloor$\\ $\text{padding[i][1]} = \text{\texttt{output\_shape}[i]} - \text{\texttt{input\_shape}[i]} $\\
$ -\lfloor \frac{\text{\texttt{output\_shape}[i]} - \text{\texttt{input\_shape}[i]}}{2} \rfloor)$\end{tabular} \\ \hline
\multirow{2}{*}{Space2Batch}                                                            & block\_size         & \texttt{output\_shape}                                                                               & $\lfloor \sqrt{\text{\texttt{output\_shape}[0]}} \rfloor$                                                                                                                                                                                                                                                 \\ \cline{2-4} 
                                                                                        & padding            & \begin{tabular}[c]{@{}c@{}}\texttt{input\_shape}\\ \texttt{output\_shape}\\ block\_size\end{tabular}            & \begin{tabular}[c]{@{}c@{}}$\text{padding[i][0]} = \lfloor \frac{\text{\texttt{output\_shape}[i+1]}*\text{block\_size} - \text{\texttt{input\_shape}[i+1]}}{2} \rfloor$\\ $\text{padding[i][1]} =  \text{\texttt{output\_shape}[i+1]}*\text{block\_size} $\\
                                                    $ -\text{\texttt{input\_shape}[i+1]} -  \text{padding[i][0]}$\end{tabular}          \\ \hline

\end{tabular}
\caption{Rules for computing the operators' attributes. The first column is the operator types; The second column is operator attributes to rebuild; The third column is the information of the corresponding operator in the original model; The last column is the rule to compute the operator's attributes with the original DL model's information. \label{tab:attr_com}}

\vspace{-0.5cm}
\end{table*}

The in-App DL model and its framework are primarily designed for efficient inference. Most on-device DL frameworks do not support gradient computation and backpropagation. However, These training-specific computations are the foundation of the white box DL model assessment.
Given that all in-App models are converted or compiled from the BP-enabled models, \sysname proposes a BP-enabled model reconstruction technique for better assessment performance. 

The model reconstruction technique takes as input an inference-only DL model and produces its corresponding BP-enabled version. It can port the DL model from the inference-only DL framework (e.g., TFLite) to a DL framework that supports model training (e.g., TensorFlow). 
Recall existing converters have poor support for converting the inference-only model to a BP-enabled one. 

DL model can be viewed as a computational graph that defines how to process the inference input. The computational graph is a directed graph to represent the DL computation procedure. The computational graph has three kinds of nodes, i.e., operator node, parameter node, and input node. Operator nodes represent the basic mathematical computation in the DL model, such as convolution, dense, padding, multiply, etc. The operator nodes are characterized by operator type and operator attributes. Parameter nodes represent the BP-enabled parameters of the neural network, e.g., weights and bias. Parameter nodes, input nodes, and operator nodes can feed their value (i.e., tensor) into other operator nodes. Operator nodes compute the output given values for its inputs. 

The model reconstruction consists of three key steps. \sysname first extracts the structure of the computational graph. The structure conveys the data dependencies between the nodes and the type of the operator nodes. Then \sysname computes the values of operator attributes and the values of the parameter nodes. Finally, \sysname generates the model code from the constructed computational graph.

\noindent \textbf{Graph Structure Extraction.} Before extracting the structure, we define a representation for the DL model's computational graph. We build a directed graph, which can represent the data dependency among nodes. The input node is characterized by input shape and output shape. The operator node is characterized by input shapes, output shapes, operator types, and operator attributes. Output shapes and values characterize the parameter node. \sysname's graph representation shares the same design with the computational graph of the widely-used Keras, in terms of operators and parameters. 

First, \sysname parses the in-App DL model according to get the information about model structures, operators, and parameters. 
Then, \sysname uses the parsed information to rebuild the structure of the computational graph with our representation. 
The rebuilt computational graph conveys the completed computation procedure, and types of operator nodes are determined.

\noindent \textbf{Node Value Computation.} After getting the graph structure, \sysname builds the values of operator attributes and the values of parameter nodes. Note that operator attributes represent the operator configuration. For example, the convolution's attributes contain the kernel size, strides, padding strategy, and output channel. The parameter nodes' values are constants used by operators (e.g., the weights of the convolution operator), and the values are determined during training. 

The parameter values in different DL frameworks are represented as tensors with different axis orders. Our representation shares the same axis order with Keras. So, we can compute the parameter values by converting the axis of the in-App model's parameter values. Some of the parameter values, e.g., weights of convolution operation, are quantified. \sysname undoes the quantification according to the quantization rules. For example, we use $v= s\times(q - z)$ to undo the TFLite's quantization, where the $v$ is the reversed weights, $q$ is the quantized value, $s$ denotes the scale coefficient, and $z$ denotes the zero point value. 

As for the attribute values, \sysname proposes a rule-based computation method by fully considering the operators' computation characteristics. We can obtain the attribute values through the DL framework-independent information. For example, when computing the attribute values of a convolution operator, the kernel size can be obtained through the output shape of the convolution operator's input parameter nodes.

Part of the rules used to compute the representative attribute values is summarized in Table~\ref{tab:attr_com}. The needed information is the information used to compute the attribute values. The needed information can be collected from the corresponding operator in the in-App model. The \texttt{input\_shape} denotes the shape of the corresponding operator's input tensor. The \texttt{output\_shape} denotes the shape of the corresponding operator's output tensor. The \texttt{weight\_shape} denotes the \texttt{output\_shape} of the corresponding operator's input parameter node.

\noindent \textbf{Model Code Generation.}
\sysname can finally generate the model code with the constructed computational graph. The generated model code is the same as that used to develop a DL model and load the trained model weights. The model code has three parts, i.e., operator initialization, flow construction, and weight loading. The operator initialization code is to utilize the operator type and attributes to create operator instances. The flow construction code is to utilize the graph structure and operator instances to construct the model's data flow. The weights loading code is to assign the trained parameters to the operator instances.

Without loss of generality, we choose the widely-used Keras as the DL framework to reconstruct the BP-enabled model. Algorithm~\ref{alg:code-gen} shows how \sysname generates the Keras model code. The algorithm takes as input the computational graph constructed before. First, \sysname adds all initialization code for every operator node (line 1$\sim$4). Then, it performs the topology sort on the computational graph to determine the execution order of the operators (line 5). Next, it generates the forward-pass code according to the operator execution order (line 6$\sim$7). \sysname adds a fragment of template code to construct the DL model instance (line 8). Finally, we add the code that loads the weights of the reconstructed model and assigns them to the corresponding operator node. (line 9$\sim$10).



\begin{algorithm}[t]
\small
\caption{Model Code Generation.}\label{alg:code-gen}
\KwData{The computational graph: $graph = [node_1, node_2, ..., node_n]$}
\KwResult{$\texttt{code}$}
\texttt{code} = [] //initialization \; 
\For{\textit{node} in graph}{
    \If{\textit{node}.nodeType == 0}{
        \texttt{code}.append(\text{genInitCode}(\textit{node}))
    }
}
sortedNodeList = \text{topologySort}(graph)\;
\For{\textit{node} in sortedNodeList}{
    \texttt{code}.append(\text{genForwardCode}(\textit{node}))
}
\texttt{code}.append(modelBuildingCode)\;
\For{\textit{node} in graph}{
    \texttt{code}.append(genWeightInitCode(\textit{node}))
}
\end{algorithm}

%% file: expr.tex
\section{Evaluation}
\label{sec:oma}

We implement the \sysname where the code reconstruction consists of 6.3K lines of Java code, and the model reconstruction consists of 1.9K lines of Python code.

In this section, we first perform a large-scale study on all found iApps.
We then perform code and model reconstruction on 100 in-App models using the top two frameworks, i.e., TFLite and TensorFlow. The selected models perform vision-related tasks, which is the most widely used kind of task in iApps~\cite{xu2019first}. Next, we use \sysname to perform the robust assessment of the reconstructed results to show the effectiveness of our \sysname. Finally, we perform three representative physical adversarial attacks to demonstrate the \sysname's security meaning.

\subsection{iApp Statistics}
\label{subsec:iApp_statis}

There are 25k APKs downloaded from five App markets, i.e., APKPure~\cite{apkpure}, 360 App Store~\cite{360_app_store}, Baidu App Store~\cite{baidu_app_store}, Xiaomi App Store~\cite{mi_app_store}, and Anzhi Market~\cite{anzhi}, in June 2021. We downloaded about 5k the most popular Apps on each market. After removing the duplicated Apps according to the iApps' MD5, there are about 15k Apps left.

We find 3,064 iApps from the downloaded APKs. And the number of in-App models is 3,845. Some of these models are repeated in different iApps because they are open source or from the same intelligent SDK, e.g., Volcengine(from ByteDance)~\cite{volcengine} or SenseTime~\cite{sensetime}. After removing duplicate models, we find 800 unique in-App models. We count these in-App models according to the DL framework in Table~\ref{tab:frameworks}.

\begin{table}[h]
\small
    \centering
\begin{tabular}{|c|l|l|}
\hline
\textbf{Category}                                                                                 & \textbf{Framework/SDK}                           & \textbf{Count}                    \\ \hline
                                                                                         & TensorFlow~\cite{tensorflow2015-whitepaper}                               & 128                       \\ \cline{2-3}
                                                                                         & TFLite~\cite{tflite}                           & 123                       \\ \cline{2-3}
                                                                                         & NCNN~\cite{ncnn}                                     & 42                        \\ \cline{2-3}
                                                                                         & Caffe~\cite{jia2014caffe}                                    & 35                        \\ \cline{2-3}
                                                                                         & MNN~\cite{alibaba2020mnn}                                      & 32                        \\ \cline{2-3}
                                                                                         & TNN~\cite{tnn}                                      & 9                         \\ \cline{2-3}
\multirow{-7}{*}{\begin{tabular}[c]{@{}c@{}}Open-source\\ framework\\ (372)\end{tabular}} & ONNX~\cite{onnx}                                     & 3                         \\ \hline
                                                                                         & Volcengine(ByteDance)~\cite{volcengine}                    & 167                       \\ \cline{2-3}
                                                                                         & SenseTime~\cite{sensetime}                                & 153                       \\ \cline{2-3}
                                                                                         & {\color[HTML]{1F2329} Kwai~\cite{kwai}}              & {\color[HTML]{1F2329} 22} \\ \cline{2-3}
                                                                                         & {\color[HTML]{1F2329} MindSpore(Huawei)~\cite{mindspore}} & {\color[HTML]{1F2329} 13} \\ \cline{2-3}
                                                                                         & Huya~\cite{huya}                                     & 7                         \\ \cline{2-3}
                                                                                         & {\color[HTML]{1F2329} Meishe~\cite{meishe}}            & {\color[HTML]{1F2329} 4}  \\ \cline{2-3}
\multirow{-7}{*}{\begin{tabular}[c]{@{}c@{}}Other SDK\\ (428)\end{tabular}}              & Unknown                                   & 62                        \\ \hline
\end{tabular}
\caption{Statistics on on-device DL inference frameworks. The last column counts the unique models using the corresponding framework. \label{tab:frameworks}}

\vspace{-0.5cm}
\end{table}

We first divide these unique models into two categories according to the DL framework. One kind of framework is the open-source DL framework, e.g., TFLite. The other kind of framework is developed for private or licensed use by commercial companies. Among all the in-App models using open-source frameworks, the models using tflite and tensorflow frameworks account for 67.5\% of the models' total number.

Of all models using private frameworks, the models utilizing frameworks provided by Volcengine and SenseTime account for 74.8\%.
We find that although the number of these models is large, these models' functionalities are similar. About 80\% of the models' functionalities are related to the feature detection of face and body, e.g., face landmark detection, pose detection, and face detection. We find that the reason for a large number of such models is that different iApps use different versions of SDKs provided by these companies, and the in-App models of the same functionalities will also be updated with the SDK.


We also count the model sizes of different DL frameworks. We find the median and mean size of in-App models are 396 KB and 1,270 KB, respectively. And about 90\% of the models are less than 3 MB in size.

\subsection{Code Reconstruction Evaluation}
\label{subsec:cre}



\begin{table}[h]
\centering
\small
\begin{tabular}{|l|l|l|l|}
\hline
\textbf{Model Task}   & \textbf{Count} & \textbf{Mean} & \textbf{Median} \\ \hline
Sytle Transformation  & 54             & 1059.9        & 530.8           \\ \hline
Classification        & 11             & 19484.5       & 8719.2          \\ \hline
Object Detection      & 11             & 8063.2        & 6737.2          \\ \hline
Super Resolution      & 8              & 2309.4        & 1974            \\ \hline
Semantic Segmentation & 7              & 6796.1        & 2714.9          \\ \hline
OCR                   & 3              & 4435.0        & 2134.8          \\ \hline
Text Detection        & 3              & 4905.3        & 2711.3          \\ \hline
Pose Estimation       & 1              & 2304.6        & 2304.6          \\ \hline
Depth Estimation      & 1              & 252.2         & 252.2           \\ \hline
Face Comparison        & 1              & 89157.7       & 89157.7         \\ \hline
\end{tabular}
\caption{Statistics on the selected 100 unique DL models by task. The second column counts the amount of models for ecah task. The third and forth column shows the mean and median of the model size (KB).}
\label{tab:tf_and_tflite_model_task_cnt}
\vspace{-0.6cm}
\end{table}

We select \textbf{\textit{100 unique}} DL models of the two widely-used DL frameworks, i.e., TensorFlow and TFLite, to evaluate the effectiveness of reconstruction techniques. We report the evaluation results of code reconstruction here. The effectiveness of the model reconstruction is evaluated through the robustness assessment in the next part.

\subsubsection{Statistical Results}
We first count the selected models by tasks in Table~\ref{tab:tf_and_tflite_model_task_cnt}. There are 10 kinds of inference tasks performed by the selected models. The largest number of models perform style transformations, which are mainly used for image editing and beautification.

We perform the \sysname's code and model reconstruction on the desktop PC. The configuration of the desktop PC is AMD Ryzen 9 5900X CPU and 128 GB DDR4 memory.
The mean and median of the reconstructed time is 97 seconds and 17 seconds.
60\% of the reconstructions can be done in 25 seconds, and 90\% of the reconstructions can be done in 223 seconds.

We perform a study on the sliced processing code and find that \textit{(1)} IO processing mainly focuses on data type conversion, e.g., image to an array, data size adjustment, data cropping, data normalization, etc. 58\% of the iApps resize the input image to less than 512$\times$512px. \textit{(2)} Almost all iApps use a group of loop statements to process the image by pixel. \textit{(3)} The mean and median LoC of the sliced code is 1,971 and 274. 60\% of the reconstructed codes' LoC is less than 308.


\subsubsection{Representative Cases}
Here we perform three case studies to show the effectiveness of the code reconstruction.

\begin{figure}[]
  \includegraphics[width=0.4\textwidth]{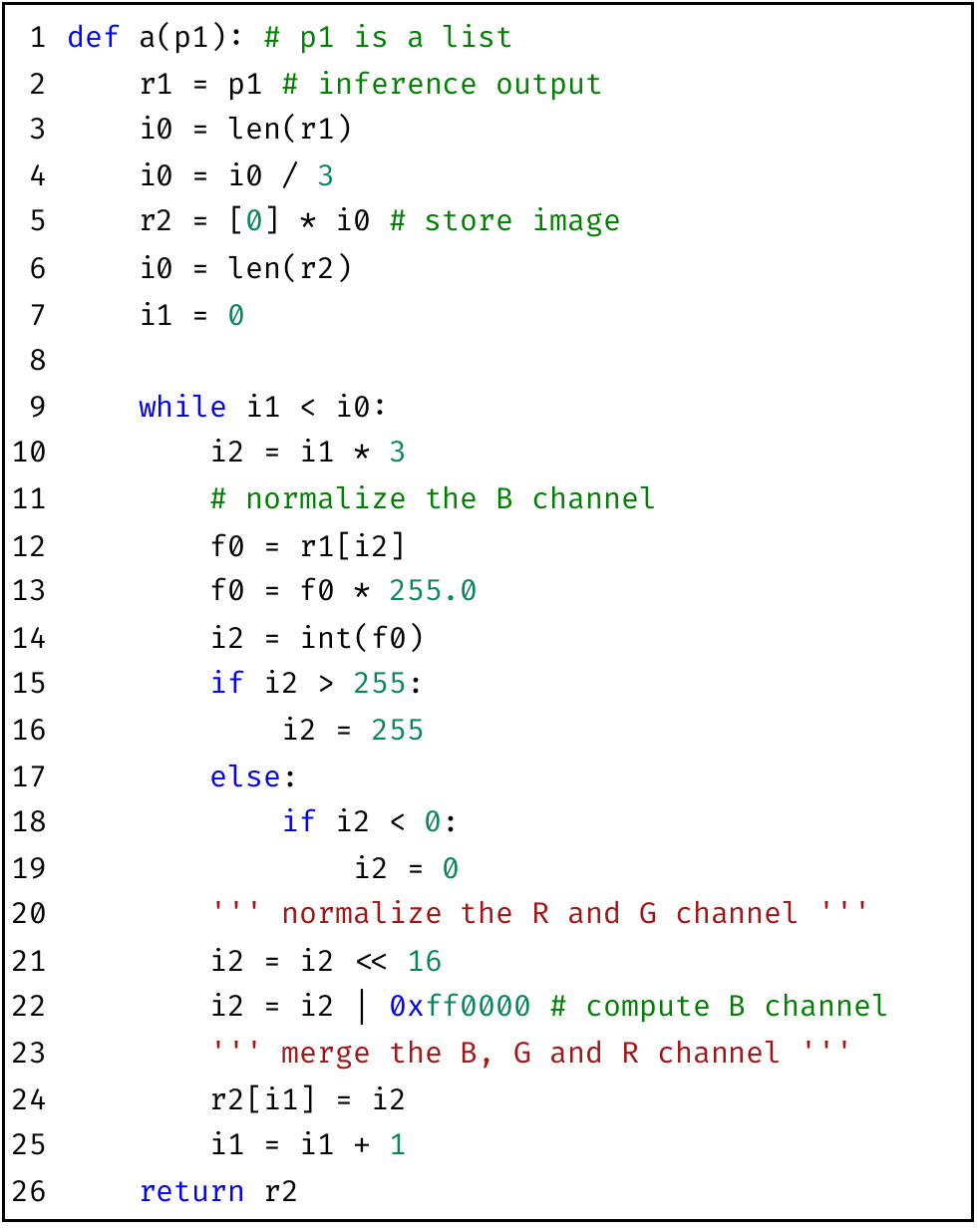}
  \caption{Part of the reconstructed code of Case 1. }
  \label{fig:case_1}
  \vspace{-0.4cm}
\end{figure}

\noindent \textit{\textbf{Case 1. Bundle ID: uk.tensorzoom.}}
This DL model is used to carry out image super-resolution. The post-processing code is responsible for converting iApp's inference result to an image (Figure~\ref{fig:case_1}). The inference result is a float array, and each float number represents each channel of each pixel (line 2). The post-processing code uses an int number (line 5) to represent a pixel with the ARGB channels, where each channel is represented by 8 bits, and the value ranges from 0 to 255. To prevent numerical overflow when converting the results, the post-processing code performs a min-max normalization (line 12$\sim$19) on the inference results.

\begin{figure}[]
  \includegraphics[width=0.4\textwidth]{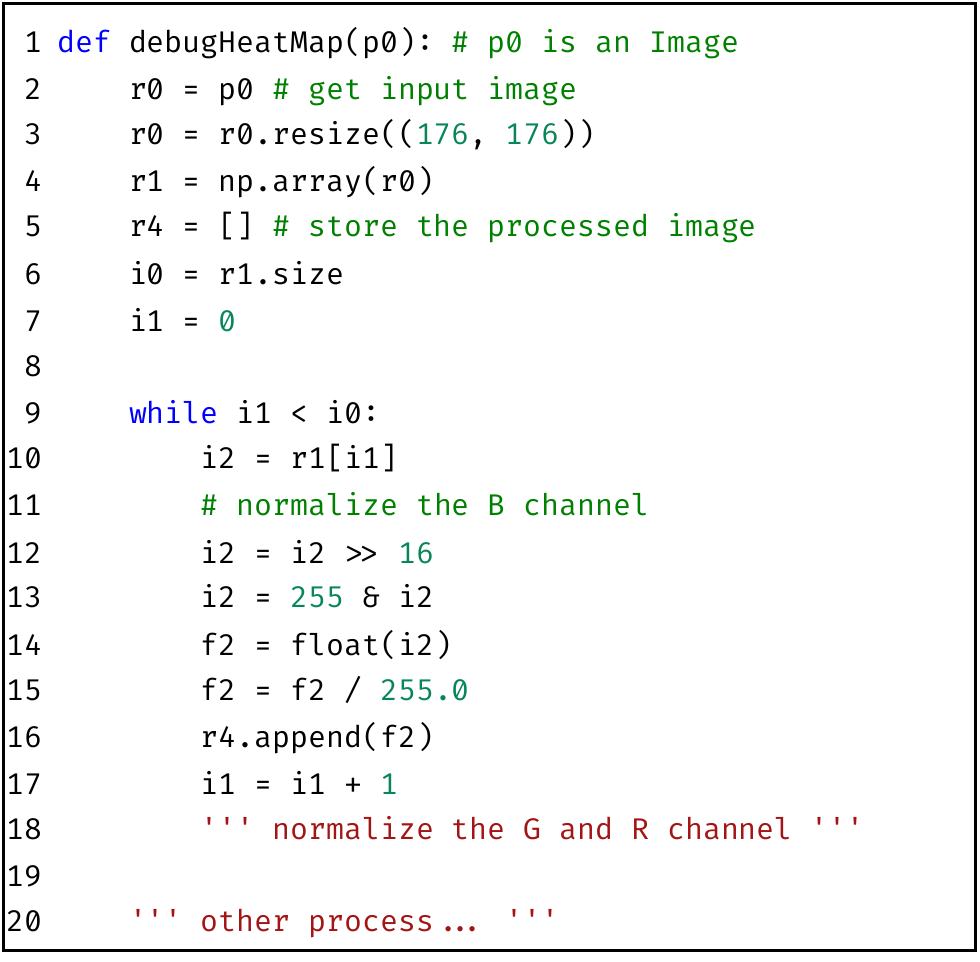}
  \caption{Part of the reconstructed code of Case 2. }
  \label{fig:case_2}
  \vspace{-0.4cm}
\end{figure}

\noindent \textit{\textbf{Case 2. Bundle ID: ru.photostrana.mobile.}}
This DL model is used to carry out face segmentation. The pre-processing code is responsible for converting the input image to a float array (Figure~\ref{fig:case_2}). This procedure involves separating the RGB channels of a pixel (line 12$\sim$13), normalizing the RGB values (line 14$\sim$15), and putting them into the array in a specific order (line 16). We find that most iApps store the input in the order of height$\times$width$\times$channel, and this iApp stores the input in the order of channel$\times$width$\times$height. If the preset channel order cannot be followed, the inference result will be wrong.

\begin{figure}[]
  \includegraphics[width=0.4\textwidth]{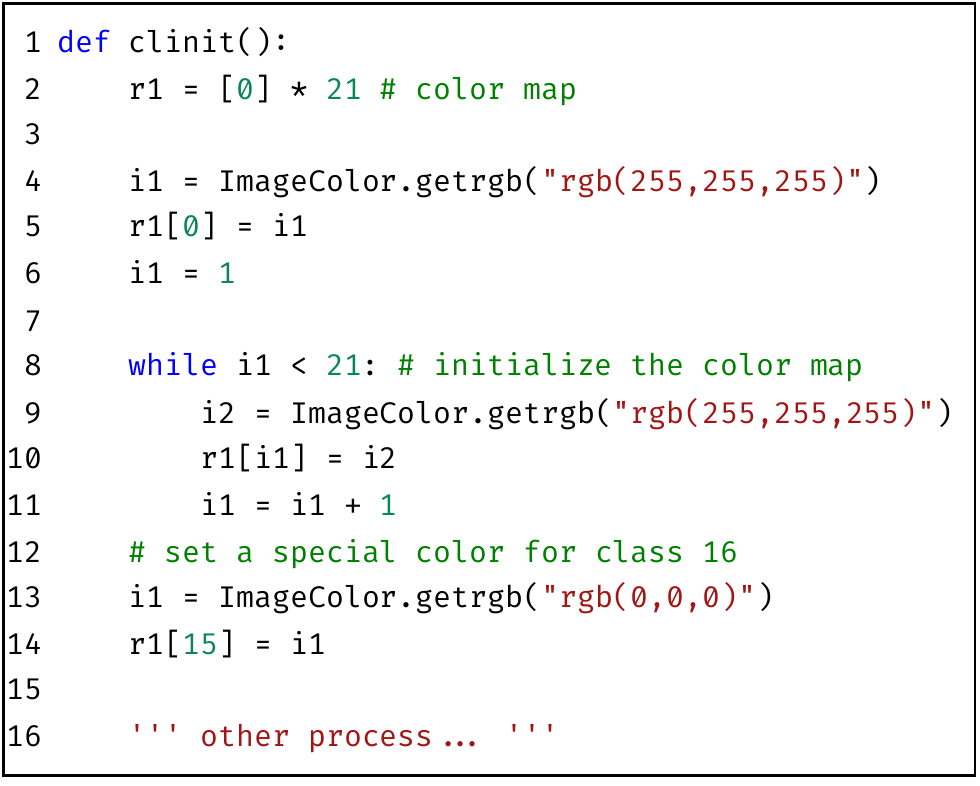}
  \caption{Part of the reconstructed code of Case 3. }
  \label{fig:case_3}
  \vspace{-0.2cm}
\end{figure}

\noindent \textit{\textbf{Case 3. Bundle ID: com.blink.academy.nomo.}}
This DL model is used to carry out a 21-class semantic segmentation. We present the post-processing code in Figure~\ref{fig:case_3}. As shown in line 13$\sim$14, the iApp only visualizes a certain class (the 15th class presents the person). Other segmentation results are discarded by the post-processing code.

\subsection{Robustness Assessment}
\label{subsec:ra}

\begin{figure*}[t]
  \includegraphics[width=0.99\textwidth]{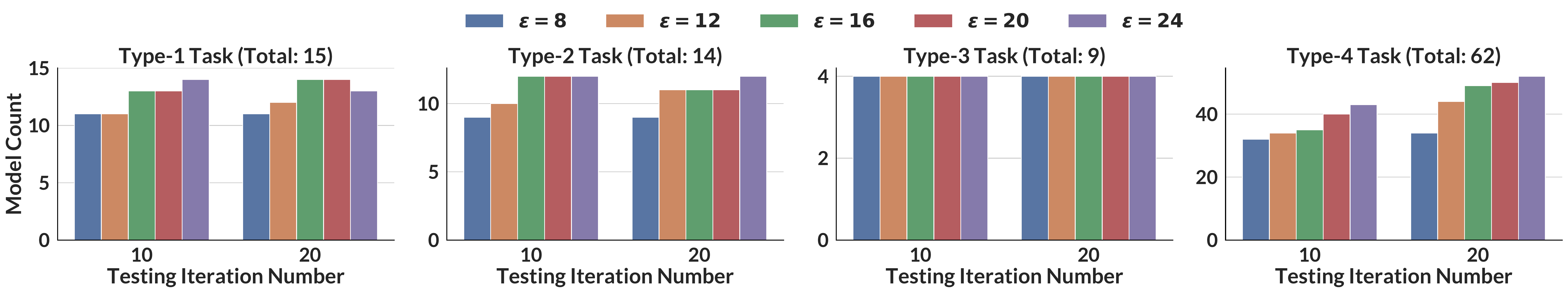}
  \caption{The robustness testing results of 100 unique in-App models by performing PGD attacks. We report the results by the type of tasks. The $\varepsilon$ is the attack budget. Testing iteration numebr represents the iteration times of the attack. The model count denotes the number of models detected to have robustness issues.}
  \label{fig:bar-pgd-res}
\end{figure*}

\begin{figure*}[t]
  \includegraphics[width=0.95\textwidth]{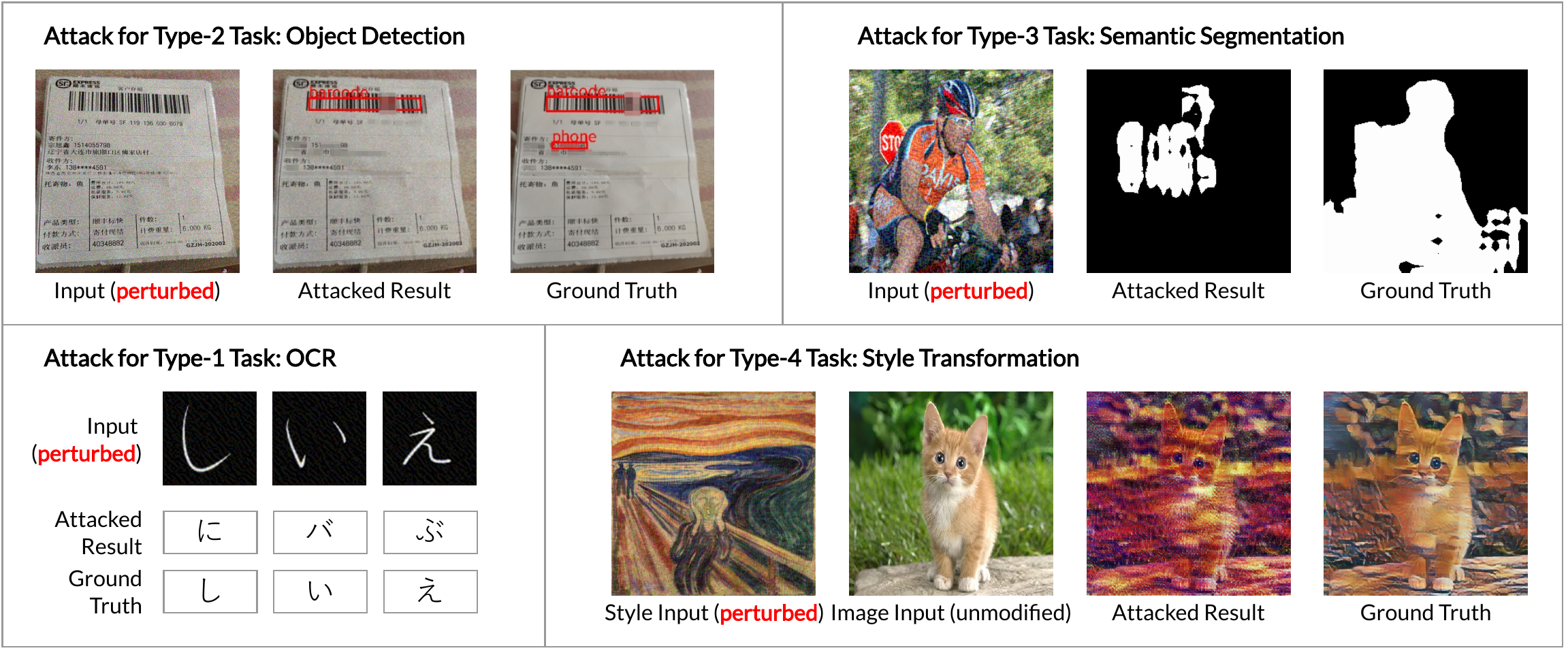}
  \caption{The robustness issues detected by \sysname.}
  \label{fig:ae_attacks}
\end{figure*}

\subsubsection{Testing Methods}
We utilize the PGD (Project Gradient Descent) attack~\cite{madry2017towards} to test the in-App model's robustness. PGD is an iterative attack method that can search for a subtle perturbation to fool the DL model. Here, we formulate the PGD attack. We denote the ground truth label of the input $x$ as $y$. We denote the adversarial example as $x_{adv}^{(t+1)}$ at $t+1$ step.

\begin{equation} \label{eq:pgd_for_classification}
    x_{adv}^{(t+1)} = \text{Proj}_{x +\mathcal{S}}\left(x_{adv}^{(t)} + \alpha \text{sgn}(\nabla_x \mathcal{L}(\mathcal{C}(x_{adv}^{(t)}),y))\right)
\end{equation}

where $\mathcal L$ is the loss function, $\mathcal C(\cdot)$ represents the model output of $x_{adv}^{(t)}$, and $\text{sgn}(\cdot)$ is the sign function. $\text{Proj}$ is a projection function which can project the input to the hypersphere $x+\mathcal S$.

Assume the original input is $x$, we can represent the final adversarial example as:
\begin{equation}
x_{adv}^{*} = x + \varepsilon
\label{eq:epsilo}
\end{equation}
Where $\varepsilon$ is the attack budget which means the manipulation limit for the input. We use MSE (Mean Square Error) as the loss function to generate the adversarial examples for different tasks.

We collect the dataset for each in-App model. Each dataset consists of 20 images. Some datasets are the subset of the open source datasets, e.g., COCO~\cite{lin2014microsoft}, VOC~\cite{Everingham10}, and CelebA~\cite{liu2015faceattributes}, and the other datasets are prepared by ourselves through the image search engine. We release sample datasets on the \textit{\textbf{anonymous website} \url{https://github.com/anonymous4896/public_data}}.

\subsubsection{Metrics}
We measure the in-App models' robustness through the following metrics for different kinds of tasks. The larger the metric value is, the higher the attack success rate is, and the worse the robustness of the in-App model. We propose four metrics according to the way of counting failure cases of different kinds of tasks.

\noindent \textbf{Type-1 Task.}
The first type of task performs classification, OCR, and face comparison.
The metric of type-1 task is
$$
\frac{1}{N}\sum_{i=1}^N 1[y_i \neq y_i']
$$
It measures the percentage of failure cases. $N$ is the number of the testing samples. The $y_i$ denotes the inferred label of $x_i$, i.e., $y_i=f(x_i)$. The $y_i'$ denotes the $x_i+\varepsilon$'s inferred label, i.e., $y_i'=f(x_i+\varepsilon)$.

In our context, \textit{if the value of the metric is larger than 0.6, we say the model is detected with robustness issues.} Note that 0.6 represents that 60\% of testing inputs are misclassified at least.

\noindent \textbf{Type-2 Task.}
The second type of task performs object detection and text detection.
For bounding boxes $B = (b_1,b_2,...,b_n)$ of all testing samples, we calculate
$$
\frac{1}{n}\sum_{i=1}^{n}1[c(b_i) \neq c(b'_i)]
$$
as the metric of type-2 task.
The $c(b)$ denotes the classification results of the bounding box $b$. The $b$ denotes the bounding box detected on the original input, and the $b'$ denotes the bounding box detected on the attacked results. Note that the value of $b_i$ may not be the same as that of $b_i'$.

In our context, \textit{if the value of the metric is larger than 0.6, we say the model is detected with robustness issues.} Note that 0.6 represents that 60\% of bounding boxes are misclassified at least.

\noindent \textbf{Type-3 Task.}
The third type of task performs semantic segmentation, depth estimation, and pose estimation.
For an image of size M$\times$N, the attack effect is measured with
$$
\frac{1}{MN}\sum_{i,j}1[s_{ij} \neq s'_{ij}]
$$
The $p_{ij}$ and $p'_{ij}$ denote the $i$-th row and $j$-th column pixel of original image and the perturbed image, respectively. The $s_{ij}$ is the semantic label of pixel $p_{ij}$, and the $s'_{ij}$ is the semantic label of the adversarial pixel $p'_{ij}$.

In our context, \textit{if the value of the metric is larger than 0.6, we say the model is detected with robustness issues.} Note that 0.6 represents that 60\% of pixels are wrongly labeled at least.

\noindent \textbf{Type-4 Task.}
The fourth kind of task performs style transformation and super resolution.
The structural similarity index (SSIM) measures the perceived quality of the attack results compared to the original results. We use
$$
1 - \texttt{SSIM}(f(x), f(x+\varepsilon))
$$
as the metric to measure the decrease of the SSIM value. The $f(\cdot)$ denotes the task. $f(x)$ and $f(x+\varepsilon)$ are the outputs of a type-4 task, when it takes as input the $x$ and $x+\varepsilon$.

In our context, \textit{if the value of the metric is larger than 0.6, we say the model is detected with robustness issues.} Note that 0.6 represents that attack makes the SSIM value drop by 0.6 at least.

\subsubsection{Testing Results}
\label{subsubsec:tr}
For each type of task, we experiment with five $\varepsilon$ (in Equation~\ref{eq:epsilo}) values and two iteration numbers. Figure~\ref{fig:bar-pgd-res} shows the testing results with different experimental settings. We count the models has robustness issues by tasks.

When performing the testing through the PDG configured with $\varepsilon$=8 and iteration number=10, robustness issues are detected in 56\% of models (56 out of 100). As the value of $\varepsilon$ increases from 8 to 16, the number of models that are detected with robustness issues increases from 56 to 78.
When we double the testing iterations (from 10 iterations to 20 iterations), we find the number of models detected with robustness issues is stable, increasing by about 14.2\%. We show some representative attack results in Figure~\ref{fig:ae_attacks}.









\subsection{Detected Physical Attacks against iApps}

\begin{figure}[]
  \includegraphics[width=0.48\textwidth]{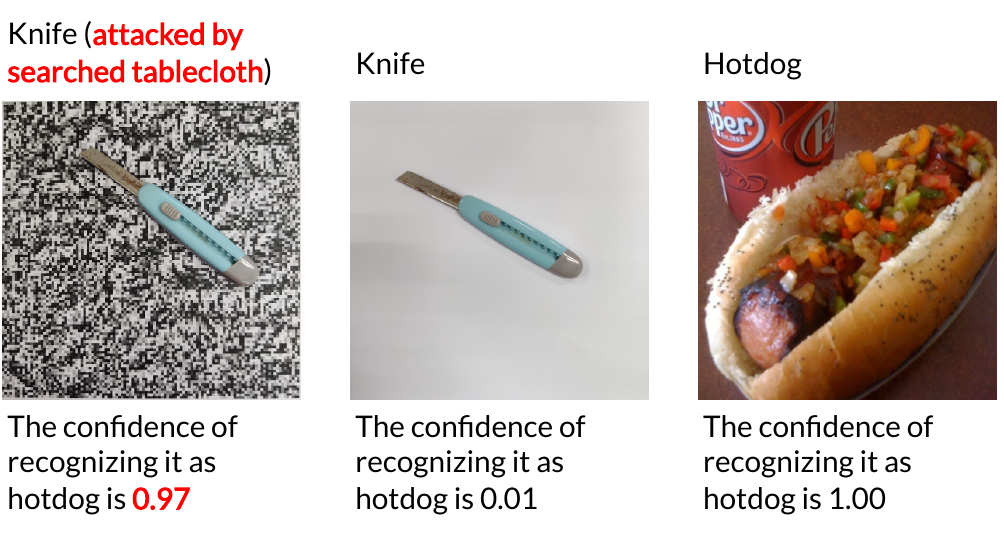}
  \caption{The physical attack results of case 1. }
  \label{fig:hot_dog}
\vspace{-0.4cm}
\end{figure}

\begin{figure*}[ht]
  \includegraphics[width=\textwidth]{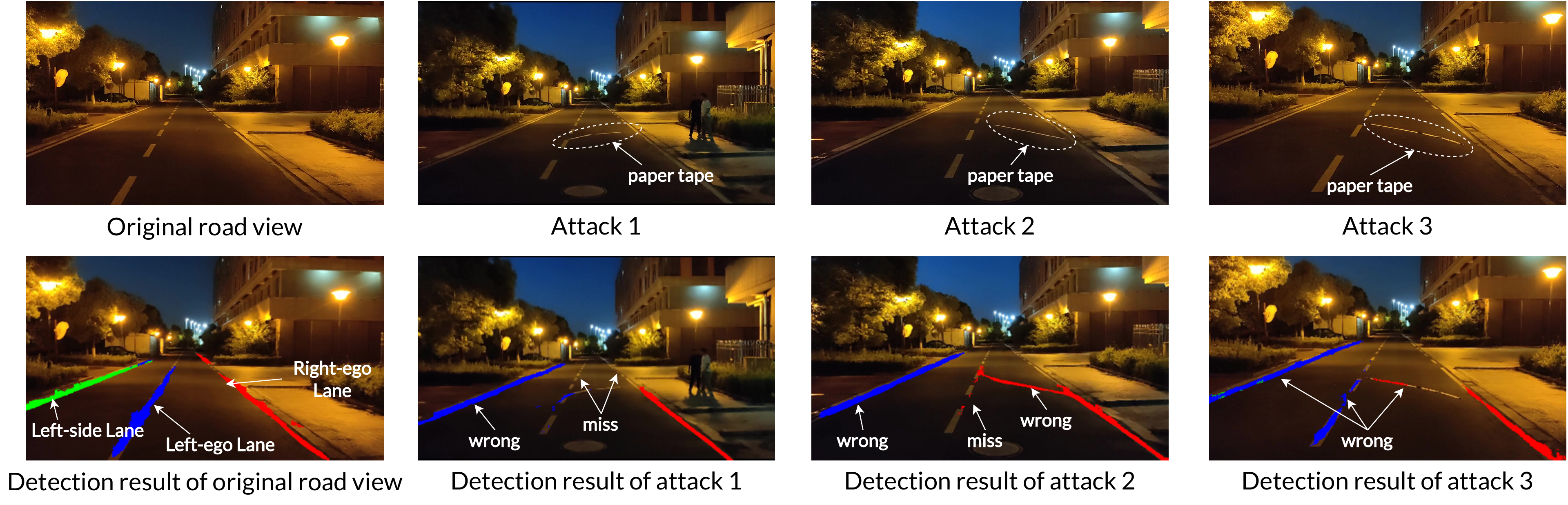}
  \caption{The physical attack results of case 2. }
  \label{fig:lane_detect}
\end{figure*}

We use \sysname to detect the security issues of three iApps by performing physical adversarial attacks.

\noindent \textit{\textbf{Case 1. Bundle ID: com.seefoodtechnologies.nothotdog.}}

This iApp can recognize whether there is a hot dog in the environment. We perform a physical attack on the in-App model. We implement that if a tablecloth with a special pattern is placed on the table, the iApp will recognize the knife on the table as a hot dog. When visually impaired people use the iApp to identify hot dogs, they face serious safety issues by mistakenly holding a knife.

We use \sysname to reconstruct the BP-enabled DL model and processing code of the hot dog detection module. Then we take a photo of the knife, segment the knife out, and generate 50 images of the knife of different sizes and rotation angles. Then we search for a specific grid background by using project gradient descent so that when we put the knife images on the grid background, the composited image is recognized as a hot dog. The loss function used to compute the gradient is the l1 norm between the hot dog's label and the recognition results of the composited image. We perform 100 rounds of the project gradient descent.

The attack results are shown in Figure~\ref{fig:hot_dog}. We show the recognition results before and after putting the knife on the searched grid tablecloth. The searched grid tablecloth makes the confidence in viewing a knife as a hot dog from 0.006 to 0.97. As a reference, the confidence of a real hot dog is 0.99.

\noindent \textit{\textbf{Case 2. Bundle ID: com.sogou.map.android.maps.}}

This iApp can be used for road navigation. By using the \sysname, we find the lane detection model in the iApp will fail when there is a paper tape in a special location on the road. The failure of lane detection can lead to serious safety problems, such as vehicles suddenly stopping or driving out of lanes.

We first use \sysname to reconstruct the BP-enabled DL model and processing code of the lane detection module. Then we choose a road and take 20 photos of the road with different camera poses. Next, we search the length and position of the paper tape, which can fail the lane detection model by using project gradient descent. The loss function used to compute the gradient is the L1 norm between the ground truth and the attack target. The attack target is to make the paper tape recognized as a lane while making the actual lane location undetectable. We perform 50 rounds of the paper tape's length and position searches.

We show the security issues detect by \sysname in Figure~\ref{fig:lane_detect}. We visualize the lane detection results on the image. The first row shows the original view and three attacks. The second row shows the corresponding attack results. The paper tape can change the lane detection results. These attacks cause the lane to be missed, or the lane's type to be incorrectly identified and brings serious security risks.

\noindent \textit{\textbf{Case 3. Bundle ID: co.mensajerosurbanos.app.mensajero.}}

This iApp can identify receipts. By using \sysname, We find that if the receipt utilizes a special background, it can evade the detection without affecting the user's reading. The iApp's competitor can cooperate with the company that prints the receipt to print the receipt with a specific background. They can degrade the receipt detection performance and cause the iApp to lose users.

\begin{figure}[ht]
  \includegraphics[width=0.48\textwidth]{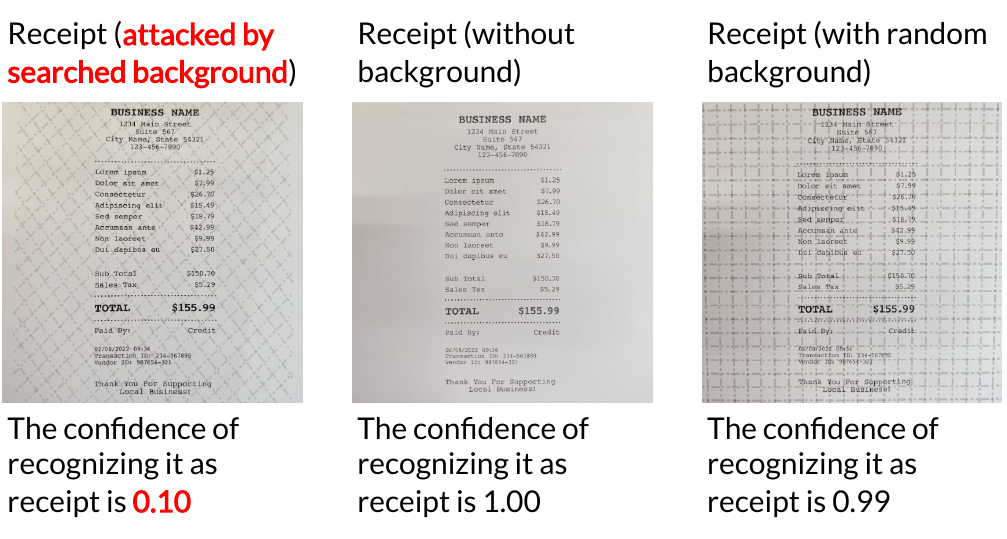}
  \caption{The physical attack results of case 3. }
  \label{fig:receipt}
\vspace{-0.4cm}
\end{figure}

We use \sysname to reconstruct the BP-enabled DL model and processing code of the receipt recognition module. We then search a background used to print the receipt, which can fool the receipt recognition. In order to reduce the impact of the receipt background on the readability of the receipt content, we use crossed dotted lines to form the background. Then we search for the number of dashed lines, the rotation angle, and the dashed line grayscale using the projected gradient descent. The loss function used to compute the gradient is the l1 norm between the receipt's label and the recognition results of the receipt with the searched background. We perform 50 rounds of the project gradient descent.

The attack results are shown in Figure~\ref{fig:receipt}. We show the receipt recognition results before and after adding the searched background. The searched background makes the confidence in the receipt recognition drop from 0.99 to 0.096. As a reference, the confidence of a receipt with random background is 0.99.

%% file: conclusion.tex
\section{Conclusion}

This work proposes \sysname to enable auto testing of iApps for App markets by two novel reconstruction techniques. The experimental results show that the \sysname can successfully reconstruct runnable IO processing code and BP-enabled DL model from commercial iApps. We perform a large-scale robustness assessment on the in-App models with the \sysname's help. \sysname also detects three representative real-world attacks against iApps. We believe \sysname can be further used to enable finding new attack surfaces lying in the coupling between code and DL models.